\documentclass[preprint, showpacs,preprintnumbers,amsmath,amssymb,nofootinbib]{revtex4}
\usepackage{mathrsfs}
\usepackage{amsmath}
\usepackage[colorlinks, linkcolor=blue, citecolor=red, urlcolor=red]{hyperref}
\usepackage{amsfonts}
\usepackage{latexsym}
\usepackage{amsfonts}
\usepackage{graphicx}
\usepackage{epsf}
\usepackage{dcolumn}
\usepackage{bm}
\newcommand{\bea}[1]{\begin{eqnarray}\label{#1}}
 \newcommand{\eea}{\end{eqnarray}}

 \def\gsim{ \lower .75ex \hbox{$\sim$} \llap{\raise .27ex \hbox{$>$}} }
 \def\lsim{ \lower .75ex \hbox{$\sim$} \llap{\raise .27ex \hbox{$<$}} }
\def\/{\over}

\begin{document}

\title{\bf  Quantum interaction between two gravitationally polarizable objects in presence of boundaries}

\author{  Hongwei Yu$^{1}$, Zhao Yang$^{1}$ and   Puxun Wu$^{1,2, }\footnote{Corresponding author at pxwu@hunnu.edu.cn}$ }
\address{ $^1$Department of Physics and Synergetic Innovation Center for Quantum Effects and Applications, Hunan Normal University, Changsha, Hunan 410081, China\\
$^2$Center for High Energy Physics, Peking University, Beijing 100080, China}
\begin{abstract}

We  investigate, in the framework of the linearized quantum gravity and the leading-order perturbation theory, the quantum correction to the classical Newtonian interaction between a pair of gravitationally polarizable objects in the presence of both Neumann and Dirichlet  boundaries.
We obtain general results for the interaction potential and  find that the  presence of a boundary always strengthens in the leading-order the interaction as compared with the case in  absence of boundaries. But different boundaries yield a different degree of strengthening.  In the limit when one partner of the pair is placed very close to the Neumann boundary,  the interaction potential is larger when the pair  is  parallel with the  boundary than when it is perpendicular to,  which is just opposite to the case  when the boundary is Dirichlet where the latter is larger than the former.  In addition, we  find that the pair-boundary separation dependence of the higher-order correction term is determined by the orientation of the pair with respect to boundary,  with the  parallel case giving a quadratic behavior and the perpendicular case a linear one. 
 

\end{abstract}


\maketitle
\section{Introduction}
The classical Newtonian theory of gravity tells us that the interaction potential of two massive objects  behaves as $r^{-1}$ with $r$ being the separation  between them. This interaction is expected to be modified  if gravity is quantized.  However, a complete study of quantum corrections to the classical Newtonian interactions requires a full theory of quantum gravity which  is elusive at the present. Even though, quantum gravity effects at the low  energies  can however be analyzed by treating the general relativity as  an effective field theory or in the framework of linearized quantum gravity. For example, by summing one-loop Feynman diagrams with off-shell gravitons, it has been found that the monopole-monopole interaction provides a quantum correction, which behaves as  $r^{-3}$, to the Newtonian  force~\cite{Donoghue}.   

A direct consequence of quantization of gravity is the appearance of  quantum vacuum fluctuations of gravitational fields, i.e., fluctuations of spacetime itself.  These  fluctuations are  
expected to induce instantaneous quadrupole  moments  in gravitationally polarizable  objects.  As a result, the induced  quadrupole-quadrupole interaction produces  a quantum correction to  the classical Newtonian interaction, which has been studied in different contexts~\cite{Ford16, Holstein17, Wu16}.  The quantum  potential between gravitational quadrupoles   is found to behave as $r^{-11}$ and $r^{-10}$ in the far and near regimes respectively.  Recently, the quadruple-quadruple interaction was extended to include  the contribution of fluctuations of thermal gravitons at finite temperature~\cite{Wu17}. In the high-temperature limit, the potential behaves  like $T/r^{-10}$, thus the thermal fluctuations of gravitons produce a dominant contribution, while in the low-temperature limit, the zero-point fluctuations dominate the interaction and the thermal fluctuations only generate a small correction.

It is well known that  field modes will be changed when boundaries are present~\cite{Khosravi91, Milton01, Weigand97},  which leads to  modifications of vacuum fluctuations.  Changes in vacuum fluctuations can produce observable effects. The  Casimir-Polder potential~\cite{Casimir} between two neutral atoms  near a perfectly conducting plate is an  example of such effects that arise from the changes of vacuum modes of electromagnetic fields~\cite{Power82, Safari, Spagnolo06}.  In the case of gravitation,  one also finds that interesting effects appear when boundaries are present, for example,  lightcone fluctuations are modified~\cite{Yu99, Yu00, Yu09, Yu03},  which leads to flight time fluctuations of a probe light signal from its source to a detector~\cite{Ford95}. 

In this paper, we shall examine the impact of  plane boundaries on the induced quadrupole-quadrupole interaction between a pair of gravitationally polarizable objects in vacuum. Our approach  is based upon the leading-order perturbation theory in the framework of  linearized quantum gravity~\cite{Yu99}, which has been used to investigate  quantum gravitational corrections in~\cite{Wu16, Wu17}.  Throughout this paper, the Latin indices run from $0$ to $3$, while the Greek letter is from $1$ to $3$. The Einstein convention is assumed for repeated indices and $\hbar=c=k_B=1$ is set. Here, $c$ is the light speed, $\hbar$ is the reduced Planck constant and $k_B$ is the Boltzmann constant.

\begin{figure}[!htb]
                \centering
                  \includegraphics[width=0.47\textwidth ]{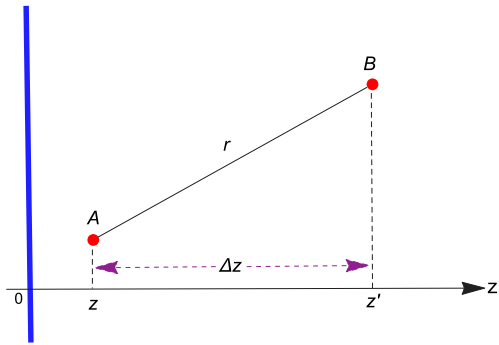}
                \caption{\label{Fig1} The system  consists of  objects $A$ and $B$  in a flat spacetime with a plane boundary at $z=0$. }
\end{figure}

\section{Basic equations} \label{secfieldequ}
The  system, which is shown in Fig.~(\ref{Fig1}), consists of two  gravitationally polarizable objects ($A$ and $B$) in a bath of fluctuating quantum  vacuum gravitational fields with a plane boundary at $z=0$.  For simplicity, we assume  $A$ and $B$  to be  described by two-level harmonic oscillators with their Hamiltonians being $H_{A(B)}=E_{A(B)}^{0} |0_{A(B)}\rangle\langle 0_{A(B)}|+E_{A(B)}^{1} |1_{A(B)}\rangle\langle 1_{A(B)}| $. For this system, the total Hamiltonian can be written as
\bea{H}
H=H_{F} + H_{A}+H_{B}+H_{AF} +H_{BF}\ ,
\eea
where $H_{F}$ is the Hamiltonian of gravitational fields  and  \bea{HA}
{H}_{A(B)F}=-\frac{1}{2}  {Q}_{A(B)}^{ij}  {E}_{ij}
\eea
 represents the interactions between the objects and gravitational fields. 
Here   $ {Q}_{ij}$ is the object's quadrupole moment   induced by the gravitational vacuum fluctuations and the gravito-electric tensor  $E_{ij}$ is defined as $E_{ij}=R_{0i0j}$ by analogy of  linearized Einstein field equation  with the Maxwell equations~\cite{Campbell}, where   $R_{\mu\nu\alpha\beta}$ is  the Riemann tensor defined in terms of the metric tensor.  
  A fluctuating  metric tensor can be expanded in a flat background spacetime as  $g_{\mu\nu}=\eta_{\mu\nu}+h_{\mu\nu}$ with $h_{\mu\nu}$ being the linearized perturbations which can be  quantized as~\cite{Yu99}
\bea{hij}
h_{ij}({\bf x}, t)=\sum_{{\bf k}, \lambda}[a_{\lambda}(\omega, {\bf x}) f^\lambda_{ij, {\bf k}}+H. c.], 
\eea
 where $H.c.$ denotes the Hermitian conjugate, ${\bf k}=\{k_1, k_2, k_3\}$, ${\bf x}=\{x, y, z\} $, $a_{\lambda}(\omega, {\bf x}) $ is the gravitational field  operator, which defines the vacuum $a_{\lambda}(\omega, {\bf x}) | \{0\}\rangle=0$, $\omega=\sqrt{k_1^2+k_2^2+k_3^2}$,  $\lambda$ labels the polarization states,  and $f^\lambda_{ij, {\bf k}} ({\bf x}, t)$ is the field mode. 
Substituting the metric tensor into the Riemann tensor gives  
\bea{Eij}
E_{ij}=\frac{1}{2}\ddot{h}_{ij}\;,
\eea
where a dot denotes a derivative with respect to time $t$.

Using the   leading-order perturbation theory, we find that  the interaction potential between  two objects, which is just the shift of the ground-state energy, arises from    fourth-order perturbations~\cite{Casimir, Buhmann, Safari} and  can be expressed as 
\bea{ES}
U_{AB}({\bf x}_{A}, {\bf x}_{B}) =&-&{\sum_{\mathrm{I,II,III}}}' \frac{\langle 0| \hat{H}_{AF}+ \hat{H}_{BF}| \mathrm{I} |\langle\mathrm{I} | \hat{H}_{AF}+ \hat{H}_{BF}| \mathrm{II} \rangle}{(E_{\mathrm{I}}-E_{0})(E_{\mathrm{II}}-E_{0})} \nonumber \\  && \qquad \times \frac { \langle\mathrm{II} | \hat{H}_{AF}+ \hat{H}_{BF}| \mathrm{III} \rangle \langle\mathrm{III} | \hat{H}_{AF}+ \hat{H}_{BF}| 0\rangle}{(E_{\mathrm{III}}-E_{0})}\;,
\eea
where $|0\rangle= |0_{A}\rangle |0_{B}\rangle|\{0\}\rangle$ is the ground state of the whole system, which is omitted in the summation as indicated by  a  prime,   and the summation includes position and frequency integrals.  $| \mathrm{I} \rangle$, $| \mathrm{II} \rangle$  and  $| \mathrm{III} \rangle$ are the intermediate states. 
In Ref.~\cite{Wu16} it has been shown  that there are ten possible combinations of intermediate states, which are listed in Table.~(\ref{Tab1}).
 Summing up all  of them, we obtain that  the interaction potential for  isotropically polarizable objects  can be expressed as 
\bea{UAB0}
U_{AB}({\bf x}_{A}, {\bf x}_{B}) &=& -\frac{1 }{4(\omega_{A}+\omega_{B})} \int_{0}^{\infty}d\omega \int_{0}^{\infty}d\omega' \;\frac{ \tilde{\alpha}_{A} \tilde{\alpha}_{B}(\omega_{A}+\omega_{B}+\omega)}{(\omega_{A}+\omega)(\omega_{B}+\omega)} \bigg(\frac{1}{\omega+\omega'}-\frac{1}{\omega-\omega'} \bigg) \nonumber \\
&&\qquad \qquad \times  G_{ijkl}(\omega,  {\bf x}_{A}, {\bf x}_{B}) G_{ijkl }(\omega',  {\bf x}_{A}, {\bf x}_{B})\ ,
\eea
where  $\omega_{A(B)}=(\omega_{A(B)}^{1}-\omega_{A(B)}^{0})$ with $\omega_{A(B)}^{1}=E_{A(B)}^{1}$ and $\omega_{A(B)}^{0}=E_{A(B)}^{0}$  represents the transition frequency  of  the object, $
\tilde{\alpha}_{A(B)} \equiv \tilde{Q}_{A(B)}^{ij}  \tilde{Q}_{A(B)}^{*ij} =  |\tilde{Q}_{A(B)}^{ij}|^{2} $ with $\tilde{Q}_{A(B)}^{ij}= \langle 0_{A(B)}| {Q}_{A(B)}^{ij} | 1_{A(B)}\rangle$ and $\tilde{Q}_{A(B)}^{*ij}= \langle 1_{A(B)}| {Q}_{A(B)}^{ij} | 0_{A(B)}\rangle$, and $ G_{ijkl}(\omega,  {\bf x}_{A}, {\bf x}_{B})$ is the two-point correlation function of gravito-electric fields
\bea{fijkl}
 G_{ijkl}(\omega,  {\bf x}_{A}, {\bf x}_{B})=\langle 0| {E}_{ij}(\omega, {\bf x}_{A}) {E}_{kl}(\omega, {\bf x}_{B}) |0\rangle \ .
 \eea

\section{ Neumann boundary condition}
Now we consider what happens to the potential  when a Neumann boundary is present.  For metric perturbations which satisfy the  Neumann boundary condition  $\partial_z f^\lambda_{ij, {\bf k}}|_{z=0}=0$,  the field mode $f^\lambda_{ij, {\bf k}}$ can be expressed as
\bea{fk1}
 f^\lambda_{ij, {\bf k}} ({\bf x}, t)=\sqrt{\frac{8\pi G}{2 \omega (2\pi)^{3} }}\big[e_{ij}({\bf k}, \lambda)e^{i ({\bf k} \cdot {\bf x}- \omega t)}+e_{ij}({\bf k}^-, \lambda)e^{i ({\bf k}^- \cdot {\bf x}- \omega t)}\big] \ ,
\eea
in the transverse tracefree (TT) gauge with $e_{ij}({\bf k}, \lambda)$ being polarization tensors. Here 
\bea{wk} \nonumber
{\bf k}^-=\{k_1, k_2, -k_3\} \ ,
\eea
 and  $G$ is the Newton's gravitational constant.

 From Eqs.~(\ref{hij}), (\ref{Eij}),  (\ref{fijkl}) and (\ref{fk1}), one finds that the two-point correlation function of $E_{ij}$ has the form
\bea{G9}
G_{ijkl}(r, {\bar r}, \Delta t) &=& \frac{1}{4}\langle 0| \ddot {h}_{ij}( {\bf x}, t) \ddot{h}_{kl}( {\bf x}', t') |0\rangle \\\nonumber   
&=& \frac{G}{8\pi^{2}}\int d^{3} {\bf k} \ {\omega^{3}} e^{i\omega \Delta t}\sum_{\lambda} \big[ e_{ij}({\bf k}, \lambda) e_{kl}({\bf k}, \lambda)  e^{i{\bf k}\cdot {\bf r}} + e_{ij}({\bf k}, \lambda) e_{kl}({\bf k}^-, \lambda)  e^{i{\bf k}\cdot  \bar{{\bf r}} } \\
&&\nonumber  \qquad\qquad  \quad +e_{ij}({\bf k}^-, \lambda) e_{kl}({\bf k}, \lambda)  e^{i{\bf k}^-\cdot {\bar {\bf r}}} + e_{ij}({\bf k}^-, \lambda) e_{kl}({\bf k}^-, \lambda)  e^{i{\bf k}^-\cdot {{\bf r}}}\big] \ .\eea
Here $r=|{\bf r}|$, $ \bar{r}=|{\bf \bar r}|$, and
\bea{}  {\bf r}=\{x-x', y-y', z-z'\}, \quad   {\bf \bar r}=\{x-x', y-y', z+z'\}. \eea
In the TT gauge,  the summation of polarization tensors gives~\cite{Yu99}
\bea{eijekl}
 \sum_{\lambda}\, e_{ij} ({{\bf k}, \lambda}) e_{kl} ({{\bf k}', \lambda})&=&\delta_{ik}\delta_{jl}
+\delta_{il}\delta_{jk}-\delta_{ij}\delta_{kl}
+\hat k_i\hat k_j \hat k'_k\hat k'_l+\hat k_i \hat k_j \delta_{kl} \nonumber\\
&&+\hat k'_k \hat k'_l \delta_{ij}-\hat k_i \hat k'_l \delta_{jk}
-\hat k_i \hat k'_k \delta_{jl}-\hat k_j \hat k'_l \delta_{ik}-\hat k_j \hat k'_k \delta_{il}\,,
\eea
where 
\begin{equation}
\hat k_i=\frac{ k_i}{ \omega}\,.
\end{equation}
From this summation of polarization tensors, we can obtain two following relations
\bea{eijekl2}
\sum_{\lambda}\, e_{ij} ({{\bf k}, \lambda}) e_{kl} ({{\bf k}, \lambda}) e^{i {\bf k}\cdot {\bf r}}&=&\frac{1}{\omega^4}[(\delta_{ik}\delta_{jl}
+\delta_{il}\delta_{jk}-\delta_{ij}\delta_{kl})\nabla^4
+ (\partial_i \partial_j \delta_{kl}+\partial_k \partial_l \delta_{ij} \\ \nonumber
&&-\partial_i \partial_l \delta_{jk}
-\partial_i \partial_k \delta_{jl}-\partial_j \partial_l \delta_{ik}-\partial_j \partial_k \delta_{il})\nabla^2 +\partial_i \partial_j \partial_k\partial_l] e^{i {\bf k}\cdot {\bf r}} \\ \nonumber 
&\equiv& \frac{1}{\omega^4}\,  \hat{g}^{r}_{ijkl}  \, e^{i {\bf k}\cdot {\bf r}} \,,
\eea
and 
\bea{eijekl3}
\sum_{\lambda}\, e_{ij} ({{\bf k}, \lambda}) e_{kl} ({{\bf k}^-, \lambda}) e^{i {\bf k}\cdot \bar{\bf r}}&=&\frac{1}{\omega^4} \sigma_{km}\sigma_{ln} [(\delta_{im}\delta_{jn}
+\delta_{in}\delta_{jm}-\delta_{ij}\delta_{mn})\nabla^4
+ (\partial_i \partial_j \delta_{mn}+\partial_m \partial_n \delta_{ij}\nonumber  \\ \nonumber
&&-\partial_i \partial_n \delta_{jm}
-\partial_i \partial_m \delta_{jn}- \partial_j \partial_n \delta_{im}- \partial_j \partial_m \delta_{in})\nabla^2 +\partial_i \partial_j \partial_m\partial_n] e^{i {\bf k}\cdot \bar{\bf r}} \\ 
&\equiv& \frac{1}{\omega^4} \, \sigma_{km}\sigma_{ln} \ \hat{g}^{\bar r}_{ijmn} \, e^{i {\bf k}\cdot \bar{\bf r}} \,,
\eea
where $\hat{g}^{r}_{ijkl}$ is a differential operator whose definition straightforwardly follows from Eq.~(\ref{eijekl2}),   $\sigma=\{\{1,0,0\}, \{0,1,0\}, \{0,0,-1\}\}$, $\nabla^2=\partial_i\partial^i$ and $\partial_i= \partial_{x_i}$. Substituting Eqs.~(\ref{eijekl2}, \ref{eijekl3}) into Eq.~(\ref{G9}) and performing the Fourier transform, one has 
\bea{G1611}
G_{ijkl}(r, {\bar r}, \omega) &=& \frac{G}{4\pi^{2}}\int d\Omega \ {\omega}  \big[ \hat{g}^{ r}_{ijkl} \, e^{i{\omega}{ r}\cos\theta} + \sigma_{km}\sigma_{ln}  \hat{g}^{ \bar r}_{ijmn} \,  e^{i{\omega}  \bar{r} \cos\theta} \big] \nonumber \\
&=& \frac{G}{\pi} \bigg[ \hat{g}^{r}_{ijkl}    \frac{\sin(\omega r)}{ r} +\sigma_{km}\sigma_{ln} \hat{g}^{\bar r}_{ijmn}  \frac{\sin(\omega \bar{r})}{ \bar{r} } \bigg]\ .\eea
where $\Omega$ is the solid angle, and the relation 
\bea{}
\int d\Omega e^{i {\bf k}\cdot  {\bf r}}= \int_0^\pi \sin\theta d\theta\int_0^{2\pi} d \phi e^{i \omega r \cos\theta}= 4 \pi   \frac{\sin(\omega r)}{\omega r}
\eea
has been used. 
Substituting Eq.~(\ref{G1611}) into Eq.~(\ref{UAB0}) gives 
\bea{UAB1711}
U_{AB}(r, \bar{r})  
&=&  -\frac{G^2 }{4 \pi^{2}(\omega_{A}+\omega_{B})}  \int_{0}^{\infty}d\omega \int_{0}^{\infty}d\omega' \;\frac{  \tilde{ \alpha}_{A} \tilde{\alpha}_{B}(\omega_{A}+\omega_{B}+\omega)  }{(\omega_{A}+\omega)(\omega_{B}+\omega)} \bigg(\frac{1}{\omega+\omega'}-\frac{1}{\omega-\omega'} \bigg)    \nonumber \\
& \times&
 \bigg( \hat{g}^{ r}_{ijkl}\frac{\sin(\omega r)}{r}+ \sigma_{km}\sigma_{ln} \hat{g}^{\bar r}_{ijmn} \frac{\sin(\omega\bar{ r})}{\bar{r}} \bigg)    \nonumber \\
& \times&
\bigg( \hat{g}^{\tilde{r}}_{ijkl} \frac{\sin(\omega' \tilde{r})}{ \tilde r}+ \sigma_{km'}\sigma_{ln'}  \hat{g}^{\tilde{\bar r}}_{ijm'n'} \frac{\sin(\omega' \tilde{\bar{ r}} )}{\tilde{\bar{ r}} } \bigg) |_{\tilde{r} \rightarrow r, \tilde{\bar{ r}}\rightarrow \bar{r}}  \ .
\eea
Defining  $y(r,  r')$ to be  
\bea{UAB3}
y(r,  r') &=& \frac{ 1 }{(\omega_{A}+\omega_{B}) } \int_{0}^{\infty}d\omega \int_{0}^{\infty}d\omega' \; \frac{ \tilde{ \alpha}_{A} \tilde{\alpha}_{B} (\omega_{A}+\omega_{B}+\omega)  }{(\omega_{A}+\omega)(\omega_{B}+\omega)}\bigg( \frac{1}{\omega+\omega'}+\frac{1}{-\omega+\omega'} \bigg) \nonumber  \\ &&\qquad \times \frac{\sin(\omega{r})}{ r} \frac{\sin(\omega' {r'})}{  r'}  \nonumber \\   \nonumber
 & =&\frac{ 1 }{(\omega_{A}+\omega_{B}) } \int_{0}^{\infty} d\omega \frac{\sin(\omega{r})}{ r} \int_{-\infty}^{\infty}d\omega' \; \frac{ \tilde{ \alpha}_{A} \tilde{\alpha}_{B} (\omega_{A}+\omega_{B}+\omega)  }{(\omega_{A}+\omega)(\omega_{B}+\omega)} \\ &&\qquad \times \bigg( \frac{1}{\omega+\omega'}+\frac{1}{-\omega+\omega'} \bigg)  \frac{e^{i \omega' {r'}}}{2i  r'}  \nonumber \\
 &=&\frac{ \pi }{(\omega_{A}+\omega_{B}) } \int_{0}^{\infty} d\omega \frac{ \tilde{ \alpha}_{A} \tilde{\alpha}_{B} (\omega_{A}+\omega_{B}+\omega)  }{(\omega_{A}+\omega)(\omega_{B}+\omega)} \frac{ \sin(\omega{r}) \cos( \omega {r'})}{r  r'} \ , \eea 
 and following an analogy with the electric polarizability of atoms~\cite{Buhmann1} to define the object's ground-state polarizability as
\bea{}
{\alpha}_{A(B)}(\omega)=\lim _{\epsilon \rightarrow0^{+}}\frac{ \tilde{\alpha}_{A(B)} \omega_{A(B)}}{\omega_{A(B)}^{2}-\omega^{2}-i \epsilon \omega} \ ,
\eea
which satisfies $Q_{ij}(\omega)= {\alpha}(\omega) E_{ij} (\omega, \bf x)$, one can obtain that   
 \bea{} y(r,  r') =\frac{ \pi }{2 }  \alpha_A(0) \alpha_B(0)  \frac{1}{r  r' (r+ r')}\ ,
\eea
when ${r'}\rightarrow r$,  and when $r\neq {r'}$
\bea{}
y(r, r') =\frac{ \pi }{2 }  \alpha_A(0) \alpha_B(0) \bigg[ \frac{1}{r  r' (r+r')}+ \frac{1}{r  r' (r- r')} \bigg]\ ,
\eea
where  the approximate static polarizability has been assumed. 
Then,  Eq.~(\ref{UAB1711}) can be re-expressed as 
\bea{}
U_{AB}(r, \bar{r})  
&=&  -\frac{G^2 }{8 \pi} \alpha_A(0) \alpha_B(0)   \bigg( \hat{g}^{ r}_{ijkl} \,\hat{g}^{ \tilde r}_{ijkl}   \,\frac{1}{r \tilde r (r+\tilde r)}  +  \sigma_{km}\sigma_{ln}\hat{g}^{r}_{ijkl}\,  \hat{g}^{\bar r}_{ijmn} \,  \frac{1}{r \bar r (r+\bar r)} \\  \nonumber &+& 
  \sigma_{km}\sigma_{ln}\ \hat{g}^{\bar r}_{ijmn} \, \hat{g}^{ r}_{ijkl}\,  \frac{1}{r \bar r (\bar r+r)} +\hat{g}^{ \bar r}_{ijkl}\,  \hat{g}^{\tilde{\bar r}}_{ijkl} \, \frac{1}{\bar{r} \tilde{\bar r} ( \bar r+\tilde{\bar r})}  \bigg) |_{\tilde{r} \rightarrow r, \tilde{\bar{ r}}\rightarrow \bar{r}} \ .
\eea
Here $\sigma_{km}\sigma_{lm}=\delta_{kl}$ has been used. 

 After lengthy calculations, one can arrive at the interaction potential 
\bea{UAB20}
U_{AB}(r, \bar{r})  
&=&  -\frac{G^2 }{4 \pi} \alpha_A(0) \alpha_B(0)   \bigg( \frac{3987}{ r^{11}}  +\frac{3987}{ \bar{r}^{11}}  + \frac{144}{r^5 \bar{r}^{5}(r+\bar{r})^9}\big[A +Br^4 \cos4\theta \\  \nonumber && +4 C r^2\cos2\theta+ 12 B r^2 \bar{r}^2  \cos2{\theta} \cos2\bar{\theta}+4 \bar{C} \bar{r}^2  \cos2\bar{\theta}+ B\bar{r}^4 \cos4 \bar{\theta}\ \big] \bigg) ,
\eea
where
\bea{AB}
A&=&9 (r^8 + 9 r^7 \bar{r} + 37 r^6 \bar{r}^2 + 93 r^5 \bar{r}^3 + 198 r^4 \bar{r}^4 + 
   93 r^3 \bar{r}^5 + 37 r^2 \bar{r}^6 + 9 r \bar{r}^7 + \bar{r}^8),  \nonumber \\
  B&=& 3 r^4 + 27 r^3 \bar{r} + 83 r^2 \bar{r}^2 + 27 r \bar{r}^3 + 3 \bar{r}^4 , \nonumber  \\
  C&=&-3 r^6 - 27 r^5 \bar{r} - 100 r^4 \bar{r}^2 - 180 r^3 \bar{r}^3 + 60 r^2 \bar{r}^4 + 
 27 r \bar{r}^5 + 3 \bar{r}^6,  \nonumber \\ 
\bar{C}&=& - 3 \bar{r}^6 -
 27 r \bar{r}^5- 100 r^2 \bar{r}^4- 180 r^3 \bar{ r}^3 +60 r^4 \bar{r}^2  + 27 r^5 \bar{r}+    3 r^6  .
   \eea
Here $\theta$ and $\bar{\theta}$ are the angles of $\bf r$ and  $\bar{\bf r}$ with respect to the normal direction of the  plane boundary, respectively.  The potential includes three terms:  the usual $r^{-11}$ interaction potential between two objects in the absence of the plane boundary~\cite{Ford16,Wu16},  the ${\bar r}^{-11}$ term which is the interaction between the object $A$ and the image of object $B$ reflected by the plane boundary, and   the remaining term depending on both $r$ and $\bar r$. 

\subsection{Two special cases}
Now we analyze the interaction potential in some special circumstances. The first special case is that two objects are placed in parallel with the plane boundary ($ z-z'= 0$), which means that $\theta=\frac{\pi}{2}$, $\bar{\theta}=\cos^{-1} \frac{2z}{\bar{r}}$ and $\bar{r}=\sqrt{r^2+4 z^2}$. When the two-object system is close to the boundary, i.e. when $z\ll r$ ($r\sim\bar r$), we find that 
\bea{}
U_{AB}(r) &=&  -\frac{G^2}{4\pi} \alpha_A(0)  \alpha_B(0) \bigg (\frac{10242}{r^{11}}-119790\frac{z^2}{r^{13}} \bigg) \ . \eea 
It is easy to see that the boundary  increases the potential about $2.6$ times in the leading-order since the coefficient in the case of flat spacetime without boundary  is $3987$ although the boundary do not change  the behavior of  $r$-dependence. The boundary also gives a negative higher-order correction term, which is dependent on $z^2$. 

Now we consider that  two objects are placed perpendicular to the boundary. Then, one has  $\theta=\bar{\theta}=0$ and $\bar{r}=r+2z$. In the limit of $z\ll r$, the potential becomes 
 \bea{}
U_{AB}(r) &=&  -\frac{G^2}{4\pi} \alpha_A(0)  \alpha_B(0)  \bigg (\frac{9252}{r^{11}}-101772\frac{z}{r^{12}} \bigg) \ . \eea 
which is, in the leading-order, about $2.3$ times that in the absence of the plane boundary, and is less than that  in the parallel case. In addition, we find that the higher-order $z-$dependent correction term is different from that in the parallel case which relies on $z^2$.

\section{ Dirichlet boundary condition}
For  the  Dirichlet boundary condition,  the field mode satisfies $f^\lambda_{ij, \bf k}|_{z=0}=0$ and thus   can be written as
\bea{fk2}
 f^\lambda_{ij, {\bf k}} ({\bf x}, t)=\sqrt{\frac{8\pi G}{2 \omega (2\pi)^{3} }}\frac{1}{i} \big[e_{ij}({\bf k}, \lambda)e^{i ({\bf k} \cdot {\bf x}- \omega t)}-e_{ij}({\bf k}^-, \lambda)e^{i ({\bf k}^- \cdot {\bf x}- \omega t)}\big] 
\eea
in the TT gauge. 
From the above equation, one can show that the two-point correlation function defined in~(\ref{fijkl}) becomes
\bea{G161}
G_{ijkl}(r, {\bar r}, \omega) &=& - \frac{G}{4\pi^{2}}\int d\Omega \ {\omega}  \big[ \hat{g}^{r}_{ijkl} \, e^{i{\omega}{ r}\cos\theta} - \sigma_{km}\sigma_{ln}\hat{g}^{ \bar r}_{ijmn} \,  e^{i{\omega}  \bar{r} \cos\theta} \big] \nonumber \\
&=& -\frac{G}{\pi} \bigg[ \hat{g}^{r}_{ijkl}    \frac{\sin(\omega r)}{ r} - \sigma_{km}\sigma_{ln}\hat{g}^{ \bar r}_{ijmn}  \frac{\sin(\omega \bar{r})}{ \bar{r} } \bigg]  \eea and then the interaction potential reads \bea{UAB25}
U_{AB}(r, \bar{r})  
&=&  -\frac{G^2 }{8 \pi} \alpha_A(0) \alpha_B(0)   \bigg( \hat{g}^{r}_{ijkl} \,\hat{g}^{\tilde r}_{ijkl}   \,\frac{1}{r \tilde r (r+\tilde r)}  -\sigma_{km}\sigma_{ln} \hat{g}^{r}_{ijkl}\,  \hat{g}^{ \bar r}_{ijmn} \,  \frac{1}{r \bar r (r+\bar r)} \\  \nonumber &-& 
\sigma_{km}\sigma_{ln}  \hat{g}^{ \bar r}_{ijmn} \, \hat{g}^{ r}_{ijkl}\,  \frac{1}{r \bar r ( \bar r+ r)} +\hat{g}^{\bar r}_{ijkl}\,  \hat{g}^{\tilde{\bar r}}_{ijkl} \, \frac{1}{\bar{r} \tilde{\bar r} ( \bar r+\tilde{\bar r})}  \bigg) |_{\tilde{r} \rightarrow r, \tilde{\bar{ r}}\rightarrow \bar{r}} .
\eea 
Following the same procedure  as in the preceding section,  we get that in the case of the Dirichlet boundary the interaction potential is  \bea{Dirifin}
U_{AB}(r, \bar{r})  
&=&  -\frac{G^2 }{4 \pi} \alpha_A(0) \alpha_B(0)   \bigg( \frac{3987}{ r^{11}}  +\frac{3987}{ \bar{r}^{11}}  - \frac{144}{r^5 \bar{r}^{5}(r+\bar{r})^9}\big[A +Br^4 \cos4\theta \\  \nonumber && +4 C r^2\cos2\theta+ 12 B r^2 \bar{r}^2  \cos2{\theta} \cos2\bar{\theta}+4 \bar{C} \bar{r}^2  \cos2\bar{\theta}+ B\bar{r}^4 \cos4 \bar{\theta}\ \big] \bigg) ,
\eea
with $A$, $B$, $C$  and $\bar{C}$ being given in Eq.~(\ref{AB}). This result is less than the one obtained in the Neumann boundary  since the third term is  subtracted in the Dirichlet boundary while it is  added in the Neumann boundary,  which indicates that different  boundary conditions lead to different  interaction potentials between two massive objects. 
 
\subsection{Two special cases}
For the special case of two objects placed in parallel with the plane boundary, we take the limit of $z\ll r$ and obtain
\bea{Diri}
U_{AB}(r) &=&   -\frac{G^2}{4\pi} \alpha_A(0)  \alpha_B(0)  \bigg (\frac{5706}{r^{11}}- 55638 \frac{z^2}{r^{13}}\bigg)\ . \eea 
Thus, a Dirichlet boundary also reinforces the interaction, but it increases only about $1.4$ time compared with the case without boundary,  which is less than that  in the case of  a  Neumann boundary.  Another noteworthy difference is that the higher-order correction term is also less than that in the Neumann boundary case.

If objects $A$ and $B$ are placed in perpendicular to the plane boundary, we obtain
\bea{Diri2}
U_{AB}(r) &=&   -\frac{G^2}{4\pi} \alpha_A(0)  \alpha_B(0) \bigg (\frac{6696}{r^{11}}- 73656 \frac{z}{r^{12}}\bigg) \eea 
in the limit of $z\ll r$, which is about $1.7$ times that in the absence of the plane boundary and is less than that from the Neumann boundary. Comparing Eqs~(\ref{Diri}) and (\ref{Diri2}) reveals that  the leading term in the potential is larger when the pair of the objects is perpendicularly placed than when it is in parallel with the boundary, which is different from the Neumann boundary case where the former is less than the latter. Similar to the Neumann boundary case, the z-dependence of the higher-order correction term in the present case is  also different from that of the parallel case. 

\section{conclusion}
In this paper, we have investigated the quantum correction to the classical Newtonian force  between a pair of polarizable objects in the presence of plane boundaries in the framework of the linearized  quantum gravity and the leading-order perturbation theory. Two kinds of boundary conditions, i.e., Neumann and Dirichlet,  are imposed. The general results are given in Eqs.~(\ref{UAB20}) and (\ref{Dirifin}).  In both cases, the potentials  consist of  three terms, i.e.,  the usual $r^{-11}$-dependent interaction potential between two objects in the absence of the plane boundary where $r$ is the separation of the two objects,  the ${\bar r}^{-11}$ term which is the interaction between the object $A$ and the image of object $B$ reflected by the plane boundary where $\bar r$ is the distance between the object $A$ and the image of object $B$, and   the term  depending on both $r$ and $\bar r$.   Different boundary conditions in general lead to different interaction potentials,  with the Neumann boundary   yielding a larger interaction than the  Dirichlet boundary.  

When one partner of the pair is placed very close to the boundary ($z\ll r$), where $z$ is the distance between the boundary and the closer partner,  we find, for both special cases, i.e., the pair is in parallel with or perpendicular to the plane boundary, that  the boundary strengthens the interaction potential as compared with the case in the absence of a boundary.  In the Neumann boundary case, the potential in the parallel case is larger than that  of the perpendicular case, which is just opposite to the Dirichlet boundary  case where the latter is larger than the former.  In addition, we  find that the sign of the higher-order correction term is negative and the pair-boundary separation dependence of the correction  is  determined by the orientation of the object pair,  with  the  parallel case and the perpendicular case  give a quadratic   and a linear correction, respectively. 

Finally, let us briefly  comment on the issue of how to realize the boundary conditions considered in this paper in some specific physical setups. It is well known that ordinary materials can hardly  reflect nor absorb  gravitational waves~\cite{Smolin}, and thus the reflection coefficient for gravitational waves will be extremely small. However, recently, there have been interesting speculations  that quantum matter such as   superconducting films  might behave like  highly reflective mirrors that realize the Dirichlet boundary condition for gravitational waves,  since the incident gravitational waves may be reflected effectively due to the so-called  Heisenberg-Coulomb effect~\cite{Minter}.  As for the Neumann boundary condition, we do not know of any specific physical setup that can realize it. So, at present, it only remains as a theoretical curiosity. 

\acknowledgments  This work was supported by the National Natural Science Foundation of China under Grants No. 11775077,  No. 11435006, No.11690034,  and No. 11375092.

\begin{table}[ht]
\begin{center}
\begin{tabular}{cllll}
\hline
 Case  & $|\mathrm I\rangle$    &
 \hspace{1ex}
 $|\mathrm{II}\rangle$   &
 \hspace{1ex}
 $\hspace{-1ex}
 |\mathrm{III}\rangle$ \\
\hline
($1$)
& $|1_A,0_B\rangle |1^{(1)}\rangle$
      &\hspace{1ex} $|0_A,0_B\rangle
        |1^{(2)},1^{(3)}\rangle$
      & \hspace{1ex}$|0_A,1_B\rangle |1^{(4)}\rangle$
       \\
($2$)
      &$|1_A,0_B\rangle |1^{(1)}\rangle$
      &\hspace{2ex}$|1_A,1_B\rangle |\{0\}\rangle$
      &\hspace{1ex}$|0_A,1_B\rangle |1^{(2)}\rangle$ \\
($3$)
      &$ |1_A,0_B\rangle |1^{(1)}\rangle$
      &\hspace{2ex}$|1_A,1_B\rangle |\{0\}\rangle$
      &\hspace{1ex}$|1_A,0_B\rangle |1^{(2)}\rangle$ \\
($4$)
      &$|1_A,0_B\rangle |1^{(1)}\rangle$
      &\hspace{2ex}$|1_A,1_B\rangle
        |1^{(2)},1^{(3)}\rangle$
      &\hspace{1ex}$|0_A,1_B\rangle |1^{(4)}\rangle$ \\
($5$)
      & $|1_A,0_B\rangle |1^{(1)}\rangle$
      &\hspace{2ex}$|1_A,1_B\rangle |1^{(2)}, 1^{(3)}\rangle$
      &\hspace{1ex}$|1_A,0_B\rangle |1^{(4)}\rangle$ \\
($6$)
      & $|0_A,1_B\rangle |1^{(1)}\rangle$
      & \hspace{2ex}$|0_A,0_B\rangle
        |1^{(2)},1^{(3)}\rangle$ 
      & \hspace{1ex}$|1_A,0_B\rangle |1^{(4)}\rangle$ \\
($7$)
      &  $|0_A,1_B\rangle |1^{(1)}\rangle$
      &\hspace{1ex} $|1_A,1_B\rangle |\{0\}\rangle$
      & \hspace{1ex}$|1_A,0_B\rangle |1^{(2)}\rangle$ \\
($8$)
      &  $|0_A,1_B\rangle |1^{(1)}\rangle$
      & \hspace{2ex}$|1_A,1_B\rangle |\{0\}\rangle$
      & \hspace{1ex}$|0_A,1_B\rangle |1^{(2)}\rangle$ \\
($9$)
      &  $|0_A,1_B\rangle |1^{(1)}\rangle$
      & \hspace{2ex}$|1_A,1_B\rangle
        |1^{(2)},1^{(3)}\rangle$
      & \hspace{1ex}$|1_A,0_B\rangle |1^{(4)}\rangle$ \\
($10$)
      &  $|0_A,1_B\rangle |1^{(1)}\rangle$
      &\hspace{2ex}$|1_A,1_B\rangle |1^{(2)}, 1^{(3)}\rangle$
      &\hspace{1ex}$|0_A,1_B\rangle |1^{(4)}\rangle$ \\
\hline
\end{tabular}
\caption{
\label{Tab1}
Ten  intermediate states contributing to the two-objects potential.}
\end{center}
\end{table}


\end{document}